\documentclass[preprint,showpacs,preprintnumbers,amsmath,amssymb]{revtex4}
\usepackage{graphicx}
\usepackage{dcolumn}
\usepackage{bm}
\usepackage{epsfig}
\begin{document}
\hfill {\bf May  2004}

\title{Exact results for `bouncing'  Gaussian wave packets}

\author{M. Belloni} \email{mabelloni@davidson.edu}
\affiliation{%
Physics Department \\
Davidson College \\
Davidson, NC 28035 USA \\
}

\author{M. A. Doncheski} \email{mad10@psu.edu}
\affiliation{%
Department of Physics\\
The Pennsylvania State University \\
Mont Alto, PA 17237 USA \\
}

\author{R. W. Robinett} \email{rick@phys.psu.edu}
\affiliation{%
Department of Physics\\
The Pennsylvania State University\\
University Park, PA 16802 USA \\
}

\date{\today}

\begin{abstract}
We consider time-dependent Gaussian wave packet solutions of the 
Schr\"{o}dinger equation (with arbitrary initial central position, $x_0$,
and momentum, $p_0$,  for an otherwise free-particle, but 
with an infinite wall at $x=0$, so-called {\it bouncing wave packets}.  
We show how difference or mirror solutions of the form 
$\psi(x,t)-\psi(-x,t)$ can, in this case,  be normalized exactly, allowing 
for the evaluation of a number of time-dependent expectation values 
and other quantities in 
closed form. For example, we calculate $\langle p^2\rangle_t$ explicitly
which illustrates  how the free-particle kinetic (and hence total) energy is
affected by the presence of the distant boundary.
We also discuss the time dependence of the expectation values of
position, $\langle x\rangle_t$,
and momentum, $\langle p\rangle_t$, and their relation to the impulsive 
force during the `collision' with the wall. Finally, the 
$x_0,p_0\rightarrow 0$ limit is shown to reduce to a special case 
of a non-standard free-particle Gaussian solution.
The addition of this example to the literature then expands 
on the relatively small number of Gaussian solutions to quantum mechanical 
problems with familiar classical analogs (free particle, uniform
acceleration, harmonic oscillator, unstable oscillator, and uniform
magnetic field) available in closed form.
\end{abstract}


\maketitle

\section{Introduction}
\label{sec:introduction}

Closed-form wave packet solutions of the time-dependent Schr\"{o}dinger 
equation are excellent exemplary models to study the time-evolution of 
quantum systems for comparison to their classical counterparts. Because 
of their special properties, Gaussian solutions are possible for a few
of the most familiar classical systems. Free-particle Gaussian wave packet
solutions are standard fare in introductory textbooks in quantum
mechanics, and this example goes back at least
to Darwin \cite{darwin}. The problem of a particle acting under a
uniform force has similar Gaussian solutions which were first derived
by Kennard \cite{kennard} and occasionally appear in undergraduate-level
presentations \cite{robinett_book}. 
Wave packet solutions for the harmonic oscillator
are discussed in some textbooks \cite{saxon} (most often using 
propagator techniques) and, with a simple change of
variables, can also be used to describe particles in unstable equilibrium
\cite{nardone}, giving rise to solutions which exhibit the expected
exponential runaway behavior. Gaussian solutions corresponding to classical
helical motion in a uniform magnetic field \cite{magnetic} have also been 
constructed. In all of these cases, the special form of the
Gaussian solutions allows for the explicit evaluation of 
time-dependent expectation values
for position ($\langle x \rangle_t$, $\langle x^2 \rangle_t$, $\Delta x_t$),
momentum ($\langle p \rangle_t$, $\langle p^2 \rangle_t$, $\Delta p_t$), 
and other quantities, for easy comparison to classical expectations.

Another simple system, also with a clear classical analog, is given by
an otherwise free particle, but restricted to the half-line
by an infinite wall at $x=0$, namely the problem defined by the 
1D potential
\begin{equation}
      V(x) = \left\{ \begin{array}{ll}
                    0 & \mbox{for $x<0$} \\
               \infty & \mbox{for $x\geq 0$}
                                \end{array}
\right.
\label{infinite_wall_potential}
\, . 
\end{equation}
In the classical case, a point particle would move freely, with constant
kinetic (and hence total) energy, exhibiting an impulsive collision at 
the wall, resulting in a discontinuous  change in momentum. 

\begin{figure}
\epsfig{file=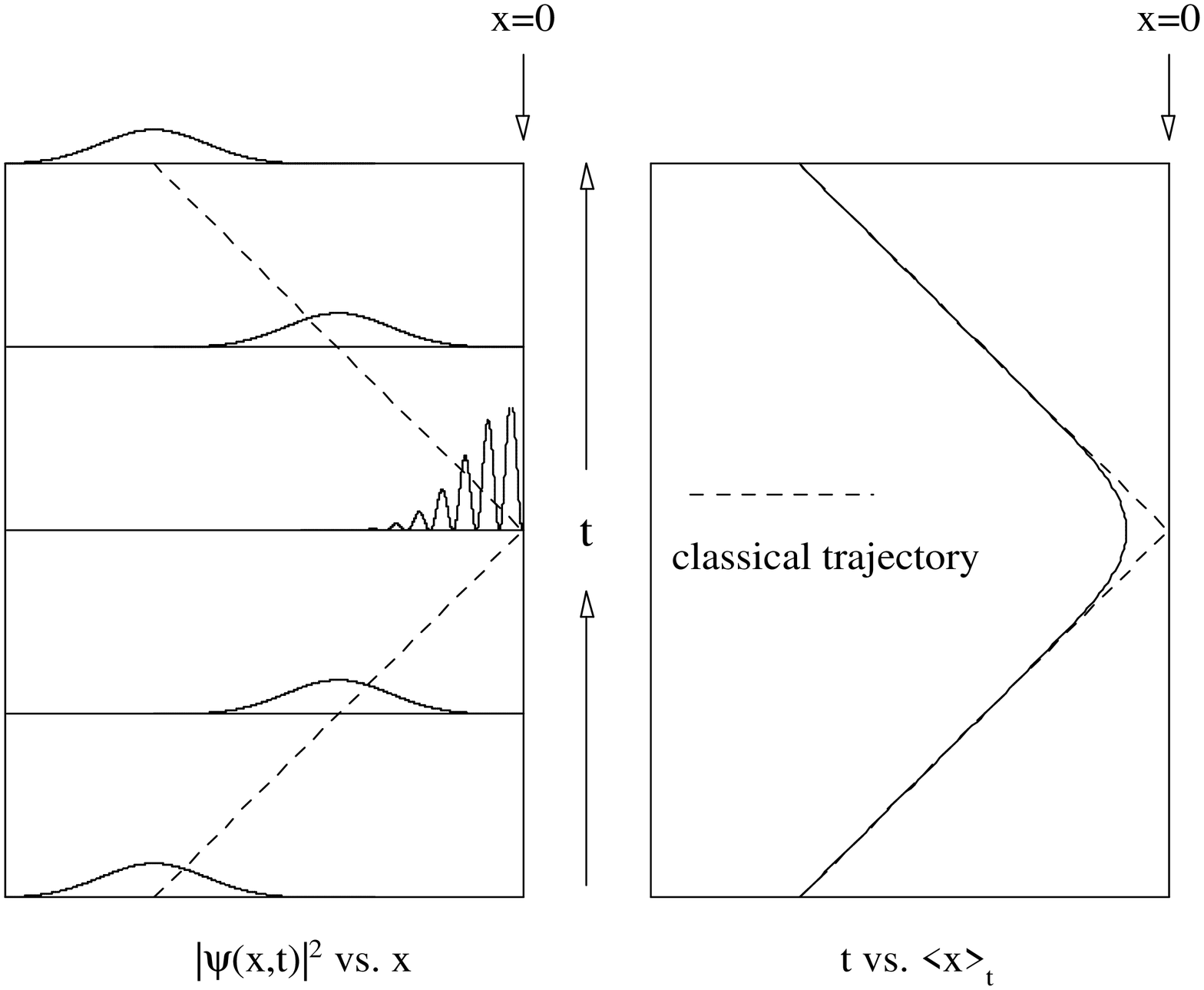,width=12cm,angle=270}
\caption{Plots of the position-space probability density,
$|\tilde{\psi}_{(G)}(x,t)|^2$ versus $x$, for the Gaussian 
{\it bouncing packet} 
of Eqn.~(\ref{andrews_solution}), for various times before, during,
and after the `collision' with the wall are shown on the left. On the
right, we show numerical calculations of the quantum mechanical
expectation value of
position, $\langle x \rangle_t$ versus $t$ (solid curve), over the same 
time range. The {\it softening} of the classical trajectory result, 
$x_{CL}(t) = -|X(t)|$ (dashed lines), near the collision in the quantum case 
is well described by Eqn.~(\ref{approx}).
\label{fig:bounce}}
\end{figure}

Localized time-dependent solutions for this problem, dubbed {\it bouncing 
wave packets}, can be constructed in a very straightforward way from
solutions of the free-particle problem. Andrews \cite{andrews} has noted 
that simple difference solutions of the
form $\psi(x,t)-\psi(-x,t)$ not only satisfy the free-particle
Schr\"{o}dinger equation for all $x$ values (if $\psi(x,t)$ does), 
but also accommodate the new boundary condition at the wall, namely 
that $\psi(0,t)=0$. This construction is very similar to image
methods in electrostatics and has been used in numerical evaluations
and visualizations of such systems \cite{thaller}, as well as for
discussions of wave packet propagation in the infinite square well
\cite{kleber}, \cite{other_mirror} (with two infinite walls producing an 
infinite series of image wave functions.)
Such solutions have also been used to visualize many aspects of the 
collision with
the wall \cite{doncheski} and we show in Fig.~\ref{fig:bounce} an example
of the time-dependent $|\psi(x,t)|^2$ using just such a construction,
illustrating the obvious interference effects between  the $\psi(x,t)$ and 
$\psi(-x,t)$ terms, near the time of the `bounce' with the wall.

The study of general relationships between expectation values, such as 
\begin{equation}
m \frac{d\langle x \rangle}{dt} =   \langle p \rangle_t
\qquad
\mbox{and}
\qquad
\frac{d\langle p\rangle_t}{dt} = 
-\left\langle \frac{dV(x)}{dx} \right\rangle_t
\label{expectation_values}
\end{equation}
is straightforward for many of the familiarly treated quantum mechanical 
systems \cite{styer}, especially for well-behaved potential energy functions.
On the other hand, for infinite well type potentials of the type considered 
here, the relationship to the classical force can be more subtle and so 
having exact or approximate closed-form solutions for quantities such as
$\langle x \rangle_t$ and $\langle p\rangle_t$ to probe such relationships 
is very useful. 
For example, the force exerted on the walls in an infinite square well
potential has been recently examined in a somewhat similar context, 
using more numerical methods, focusing on the time-dependent relationships
in Eqn.~(\ref{expectation_values}) as related to wave packet 
revivals and fractional revivals \cite{mork}.

In this note, we will show that if one uses standard Gaussian solutions
in this mirror or difference solution approach for the potential in 
Eqn.~(\ref{infinite_wall_potential}), one can also perform many 
(but not all) of the
standard calculations including exact normalization of the wave packet
solution, evaluation of many expectation values, and the calculation of
the autocorrelation function, $A(t)$, to obtain exact closed-form results,
which is the main thrust of Sec.~\ref{sec:solutions}.
This case then adds another example to the otherwise rather small pantheon
of closed-form quantum mechanical  Gaussian solutions to one-dimensional
problems with classical analogs. Using these solutions, we will also be
able to discuss the nature of the collision with the wall, deriving
approximate expressions for the impulsive force exerted on the particle 
during the collision. We will also be able to explicitly evaluate the
change in kinetic (and hence total) energy induced by the addition of the
infinite barrier, making connections to earlier work on the {\it effects
of distant boundaries} \cite{distant_boundaries} on the energy spectrum 
of quantum mechanical systems; this example extends those results to an
exact time-dependent wave packet solution and is also discussed in
Sec.~\ref{sec:solutions}.  Finally, the mirror solutions
discussed here can be connected to special, non-standard Gaussian
free-particle wave packets in the special limit when $x_0,p_0 \rightarrow
0$, as shown in Sec.~\ref{sec:special_case}.

\section{``Bouncing'' Gaussian wave packet solutions}
\label{sec:solutions}

We first review well-known textbook results for the standard 
time-dependent Gaussian free-particle momentum- and position-space 
wave packet solutions, for arbitrary initial $x_0$ and $p_0$.
These can be written in the forms
\begin{eqnarray}
\phi_{(G)}(p,t)
& = &
\sqrt{\frac{\alpha}{\sqrt{\pi}}}
\, e^{-\alpha^2(p-p_0)^2/2}
\, e^{-ipx_0/\hbar}
\, e^{-ip^2t/2m\hbar}
\label{p_gaussian_t} \\
\psi_{(G)}(x,t) & = &  \frac{1}{\sqrt{\sqrt{\pi} \alpha \hbar (1+it/t_0)}}
\,
e^{ip_0(x-x_0)/\hbar}
\, e^{-ip_0^2t/2m\hbar}
\,
e^{-(x-x_0-p_{0}t/m)^2/2(\alpha \hbar)^2(1+it/t_0)}
\label{x_gaussian_t}
\, . 
\end{eqnarray}
These solutions are characterized by
\begin{eqnarray}
\langle p \rangle_t = p_0,
& \quad  \quad&
\Delta p_t = \frac{1}{\alpha \sqrt{2}} \\
\langle x \rangle_t = x_0 + (p_0/m)t \equiv X(t),
& \quad  \quad &
\Delta x_t = \frac{\beta}{\sqrt{2}} \sqrt{1 + (t/t_0)^2} 
\equiv  \frac{\beta_t}{\sqrt{2}}
\label{standard_expectations}
\end{eqnarray}
where $t_0 \equiv m\hbar \alpha^2$ is the spreading time and
$\beta \equiv \alpha \hbar$. This gives the familiar uncertainty 
principle product
\begin{equation}
\Delta x_t \cdot \Delta p_t = \frac{\hbar}{2}\sqrt{1+(t/t_0)^2}
\quad
\longrightarrow 
\quad
\left(\frac{\hbar}{2}\right)\left(\frac{t}{t_0}\right)
\label{basic_uncertainty_product}
\end{equation}
for $t>>t_0$.

For Gaussian wave packet solutions of this type, the corresponding 
difference or mirror solution of Andrews \cite{andrews} is written as
\begin{equation}
\tilde{\psi}_{(G)}(x,t) = \left\{ \begin{array}{ll}
 N\left[\psi_{(G)}(x,t) - \psi_{(G)}(-x,t)\right] & \mbox{for $x<0$} \\
               0                              & \mbox{for $x\geq 0$}
                                \end{array}
\right.
\label{andrews_solution}
\end{equation}
where $N$ is a normalization constant. For initial free-particle
wave packets, $\psi_{(G)}(x,t)$,  which  are already 
normalized correctly (over all space), and which are sufficiently far 
apart in phase space ($x_0,p_0$ not too small), one would expect that 
$N$ would be very close (exponentially so) to unity, a result which we 
will confirm by explicit calculation below. Once normalized, of course,
any solution of the form Eqn.~(\ref{andrews_solution}) will remain
normalized for later times.

In order to evaluate the various required integrals involved in the
normalization of $\tilde{\psi}_{(G)}(x,t)$, we make use of the fact
that the difference or mirror solution is antisymmetric in $x$, so that
$|\tilde{\psi}_{(G)}(x,t)|^2$ is automatically symmetric in $x$. This 
allows us, for example,  to determine the normalization constant $N$ 
using the requirement
\begin{eqnarray}
1 & = & \int_{-\infty}^{0}\, |\tilde{\psi}_{(G)}(x),t)|^2\,dx \nonumber \\
& = &   N^2\int_{-\infty}^{0}\, |\psi_{(G)}(x,t)-\psi_{(G)}(-x,t)|^2\,dx  \\
\label{bouncing_solution}
& = &   \frac{N^2}{2} 
\int_{-\infty}^{+\infty}\, |\psi_{(G)}(x,t)-\psi_{(G)}(-x,t)|^2\,dx \nonumber
\end{eqnarray}
and because the integration region can be extended (by symmetry) over all 
space, one can do the resulting integrals in closed form to obtain
\begin{equation}
N = \frac{1}{\sqrt{1-\exp[-(x_0/\beta)^2 - (p_0\beta/\hbar)^2]}}
= \frac{1}{\sqrt{1-e^{-z_0}}}
\end{equation}
where
\begin{equation}
z_0 \equiv  \left(\frac{x_0}{\beta}\right)^2 
+ \left(\frac{p_0 \beta}{\hbar}\right)^2
= \frac{1}{2}
\left[
\left(\frac{x_0}{\Delta x_0}\right)^2
+
\left(\frac{p_0}{\Delta p_0}\right)^2
\right]
\, .
\label{z_zero}
\end{equation}
The integrals are, of course, done most simply for $t=0$, but it is then 
easy to confirm that one obtains the same answer for $t>0$, so that the 
wave function can be explicitly shown to remain normalized at all later times.
While this expression is consistent with expectations for large values of 
$x_0,p_0$, we stress that it is exact for arbitrary initial parameter 
values, even in the limit where $x_0,p_0 \rightarrow 0$,  which is discussed 
in Sec.~\ref{sec:special_case}.

The same method can be used to evaluate various expectation values
involving even powers of variables, such as $\langle x^2 \rangle_t$
and $\langle p \rangle_t^2$. For example, one can show that
\begin{eqnarray}
\langle x^2 \rangle_t & \equiv & \int_{-\infty}^{0} x^2 \, 
|\tilde{\psi}_{(G)}(x,t)|^2\,dx \nonumber \\
& = & \frac{1}{2} \int_{-\infty}^{+\infty} x^2 \, 
|\tilde{\psi}_{(G)}(x,t)|^2\,dx 
\label{x_squared} \\ 
& = & \left[
\left(x_0 + \frac{p_0t}{m}\right)^2 + \frac{\beta_t^2}{2}\right]
 + \beta_t^2 F(z_0) \nonumber 
\end{eqnarray}
where
\begin{equation}
F(z_{0}) \equiv \frac{z_{0} e^{-z_{0}}}{1- e^{-z_{0}}}
\end{equation}
and $z_0$ is defined in Eqn.~(\ref{z_zero}). 
The corresponding result for momentum is
\begin{equation}
\langle p^2 \rangle_t = 
\langle p^2 \rangle_0 = \left[p_0^2 + \frac{\hbar^2}{2\beta^2}\right]
+ \frac{\hbar^2}{\beta^2} F(z_0)
\,.
\label{p_squared}
\end{equation}
We note the similarity in form between the expressions in 
Eqns.~(\ref{x_squared}) and (\ref{p_squared}), so that the effect is
the largest (not surprisingly) in the limit when $z_0 = 0$, namely
when $x_0,p_0\rightarrow 0$. This special case also turns out to be a 
particular limit of a less-familiar free-particle solution which we 
discuss further in Sec.~\ref{sec:special_case}.

The result in Eqn.~(\ref{p_squared}) is also useful as it implies 
that the change in kinetic
(and hence total) energy caused by the addition of the distant boundary
is given by
\begin{equation}
\frac{\Delta E}{E} \approx  \frac{2F(z_0)}{1+ 2(p_0 \beta/\hbar)^2}
\, ,
\end{equation}
that is, an exponential suppression, which is consistent with more general 
arguments \cite{distant_boundaries}.

For expectation values of odd powers of $x,p$, the resulting integrals
can be done in terms of error functions (erf$(\zeta)$), but simple expressions
for special cases of interest are perhaps more useful. For example, 
the classical prediction for the position is $x_{CL}(t) = -|X(t)|$,
where $X(t) \equiv  (x_0 +p_0t/m)$; the time of the classical
collision with the wall, $t_{c}$, is determined by the condition 
$X(t_{c})=0$ and we can expand the integrals required for an evaluation of
$\langle x \rangle_{t}$ in terms of $X(t \approx t_{c})$, treated as a 
small parameter, to obtain the approximation
\begin{equation}
\langle x \rangle_{t\approx t_{c}}
= 
\int_{-\infty}^{0} x |\tilde{\psi}_{(G)}(x,t)|^2\,dx 
\approx 
\left(- \frac{\beta_t}{\sqrt{\pi}} - \frac{[X(t)]^2}{\beta_t \sqrt{\pi}}
+ \cdots\right)_{t \approx t_c}
\, .
\label{approx}
\end{equation}
The first term in this expansion was derived in Ref.~\cite{doncheski}.
We show, in Fig.~\ref{fig:bounce}, the numerically calculated value of
$\langle x \rangle_t$ versus $t$ for a sample solution and note how
the expression in Eqn.~(\ref{approx}) reproduces the softened `parabolic' 
shape of the curve near the collision time, to be compared to the purely 
classical result, namely $x_{CL}(t) = -|X(t)|$, shown as the dashed lines.

The result in Eqn.~(\ref{approx}) can then be differentiated and evaluated
at the collision time $t_{c}$, to give
\begin{equation}
\langle p \rangle_{t\approx t_{c}} = 
\left. m \frac{d\langle x\rangle_t}{dt} \right|_{t \approx t_c}
= 
- \frac{\hbar}{\beta\sqrt{\pi}}
\left[\frac{t_{c}/t_0}{\sqrt{1+(t_c/t_0)^2}}\right]
\approx 
- \frac{\hbar}{\beta\sqrt{\pi}} = - \frac{1}{\sqrt{\pi}\alpha}
\label{collision_time_momentum}
\end{equation}
for $t_{c}>>t_0$. This implies that the expectation value of momentum at the 
classical collision time does not vanish, for reasons which are most easily
visualized using the behavior of the time-dependent momentum-space
probability densities, as in Fig.~3 of Ref.~\cite{doncheski}. 
For times long before
and long after the `bounce', the momentum distribution is peaked at
$+p_0$ (within $\pm \Delta p_0$) and $-p_0$ (within $\pm \Delta p_0$)
 respectively. Roughly speaking,
at the classical collision time, the momentum components corresponding to
the fastest speeds (in the $+p_0 + \Delta p_0$ half of the distribution) 
have already been reflected from the wall and are already flipped in sign 
to have values $-(p_0+\Delta p_0)$, while
the slower components are still predominantly in the lower  $+p_0-\Delta p_0$
half of the distribution, but still positive.  
There is a small resulting asymmetry in the momentum-distribution, 
giving an expectation value of order $-\Delta p_0$,
 as in Eqn.~(\ref{collision_time_momentum}).

The result in Eqn.~(\ref{approx}) is also useful in that it gives similarly 
valid approximations for higher derivatives, such as
\begin{equation}
m \frac{d^2 \langle x \rangle_{t \approx t_c}}{dt^2}
\approx - \frac{2}{\sqrt{\pi}} \left(\frac{p_0^2}{m \beta_t}\right)
\end{equation}
which can then be used to describe the effective force exerted on the 
particle by the wall near the classical collision time. It is not clear
how one would obtain this result more directly from the Ehrenfest theorem
approach in Eqn.~(\ref{expectation_values}) using the potential
energy function in Eqn.~(\ref{infinite_wall_potential}). The dimensional 
dependence of this result on $p_0,m,\beta_t$ can be easily understood from 
simply assuming that the change in momentum during the collision is of order 
$\Delta p = p_f - p_i \approx -2p_0$, while the collision time, $\Delta t$, 
is determined by 
\begin{equation}
\frac{p_0}{m} \sim v_0 \sim \frac{\Delta x}{\Delta t}
\qquad
\mbox{or}
\qquad
\Delta t \sim \frac{\beta_t m}{p_0}
\qquad
\mbox{giving}
\qquad
F \sim \frac{\Delta p}{\Delta t} \sim -\frac{2p_0^2}{m\beta_t}
\, , 
\end{equation}
assuming we identify $\Delta x \sim \beta_t$ near the time of the collision

Finally, another useful quantity which can be evaluated in closed
form for the solution in  Eqn.~(\ref{andrews_solution}) is the
autocorrelation function, which measures the overlap between the
initial quantum state, $\psi(x,0)$, and its time-evolved value at
later times, $\psi(x,t)$. This is defined most generally by
\begin{equation}
A(t) \equiv \int_{-\infty}^{+\infty} [\psi(x,0)]^*\, \psi(x,t)\,dx
= \int_{-\infty}^{+\infty} [\phi(p,0)]^*\, \phi(p,t)\,dp
\, . 
\end{equation}
For the free-particle Gaussian solutions in Eqns.~(\ref{p_gaussian_t}) 
and (\ref{x_gaussian_t}) this can be written in the equivalent forms
\cite{robinett}
\begin{eqnarray}
A_{(G)}(t) & = &  \frac{1}{\sqrt{1-it/2t_0}} e^{ip_0^2t/2m\hbar}
\exp\left[- \frac{(X(t)-x_0)^2}{4\beta^2 (1-it/2t_0)}\right] \\ 
& = &  
\frac{1}{\sqrt{1-it/2t_0}}
\exp\left[\frac{i\alpha^2 p_0^2t}{2t_0(1-it/2t_0)}\right]
\, .
\end{eqnarray}
Both of these forms give
\begin{equation}
|A_{(G)}(t)|^2 = \frac{1}{\sqrt{1+(t/2t_0)^2}}
\exp\left[-2\alpha^2p_0^2 \frac{(t/2t_0)^2}{(1+(t/2t_0)^2)}\right]
\, .
\end{equation}
One sees that the free-particle autocorrelation function decreases 
monotonically with time, due to both the {\it dynamic} exponential 
dependence on $p_0$, as well as to the {\it dispersive} pre-factor 
which can be attributed to wave packet spreading.

For the bouncing wavepacket solution, we must evaluate the autocorrelation
using the closed form expression for the position-space solution in
Eqn.~(\ref{andrews_solution}), giving
\begin{eqnarray}
\tilde{A}_{(G)}(t) & = &  \int_{-\infty}^{0} [\tilde{\psi}_{(G)}(x,0)]^*
\, \tilde{\psi}_{G)}(x,t)\,dx \nonumber \\
& = &  N^2\int_{-\infty}^{0} [\psi_{(G)}(x,0)-\psi_{(G)}(-x,0)]^{*}
\,
[\psi_{(G)}(x,t)-\psi_{(G)}(-x,t)]\,dx
\end{eqnarray}
and the same trick of extending the integral over all space (because the
integrand is still odd under $x \rightarrow -x$) can be used to find
\begin{eqnarray}
\tilde{A}_{(G)}(t) & =&  A_{(G)}(t) 
\left(
\frac{1-\exp\left[-\left\{(x_0/\beta)^2 + (p_0 \beta/\hbar)^2\right\}/(1+it/2t_0)\right]}
{1-\exp\left[-{x_0/\beta)^2 + (p_0\beta/\hbar)^2}\right]}
\right) \nonumber \\
& = & A_{(G)}(t) \left(\frac{1-\exp[-z_{0}/(1+it/2t_0)]}{1-\exp[-z_{0}]}\right)
\, .
\end{eqnarray}
Once again, there is a monotonic decrease in $|A(t)|$ with no distinction
between the smoother time-evolution before and after the collision and the
time during the impulsive {\it splash} at the wall.

\section{Other free-particle Gaussian solutions related to the 
bouncing wave packet}
\label{sec:special_case}

While the standard Gaussian solutions for the free-particle case given
in Eqns.~(\ref{p_gaussian_t}) and (\ref{x_gaussian_t}) are the most 
familiar examples found in textbooks, it is straightforward to construct 
other localized Gaussian-like wave packet solutions, using the fact that
a wide variety of Gaussian integrals can be performed in closed form. 
Some of these can then be easily used as special case solutions for the
`bouncing' wavepacket case as well.

For example, a modified free-particle momentum-space solution of the form
\begin{equation}
\phi_{(G')}(p,t) 
= \sqrt{\frac{2\alpha^3}{\sqrt{\pi}}}
\, (p-p_0)\, e^{-\alpha^2 (p-p_0)^2/2}\, 
e^{-ipx_0/\hbar}
\,
e^{-ip^2t/2m\hbar}
\label{alt_p_gaussian_t}
\end{equation}
gives the expectation values
\begin{equation}
\langle p \rangle_t = p_0,
\qquad
\langle p^2 \rangle_t = p_0^2 + \frac{3}{2\alpha^2},
\qquad
\mbox{and}
\qquad
\Delta p_t = \sqrt{\frac{3}{2\alpha^2}}
\label{higher_p}
\end{equation}
and can be Fourier transformed to yield the position-space wavefunction
\begin{equation}
\psi_{(G')}(x,t) = i \sqrt{\frac{2}{\beta^3(1+it/t_0)^3}}
\, e^{ip_0(x-x_0)/\hbar}
\, e^{-ip_0^2t/2m\hbar}
\, (x-X(t))
\, e^{-(x-X(t))^2/2\beta^2(1+it/t_0)}
\label{alt_x_gaussian_t}
\end{equation}
with
\begin{equation}
\langle x\rangle_t = X(t)
\equiv  x_0 + \frac{p_0t}{m}
\qquad
\mbox{and}
\qquad
\Delta x_t = \sqrt{\frac{3}{2}} \beta_t
\, . 
\end{equation}
The uncertainty principle product in this case is given by
\begin{equation}
\Delta x_t \cdot \Delta p_t = \frac{3\hbar}{2} \sqrt{1+(t/t_0)^2}
\end{equation}
which is similar to that for the first excited state of the harmonic 
oscillator, at least for $t=0$. In the same way, initial 
momentum-distributions with higher 
powers of $(p-p_0)$ can be used to exhibit localized position-space 
wave packets.

For the case described by Eqns.~(\ref{alt_p_gaussian_t}) and 
(\ref{alt_x_gaussian_t}),
if we consider the special case of $x_0,p_0\rightarrow 0$, we find
the solution
\begin{equation}
\psi_{(0)}(x,t) = i \sqrt{\frac{2}{\sqrt{\pi}\beta^3 (1+it/t_0)^3}}
\, x\, e^{-x^2/2\beta^2(1+it/t_0)}
\end{equation}
which is valid for all space. This clearly satisfies $\psi(0,t)=0$ for all
$t$ and so can be used as a solution for the `bouncing' packet case
corresponding to the potential in Eqn.~(\ref{infinite_wall_potential});
the solution must then be `renormalized' by multiplying  by a factor of 
$\sqrt{2}$ to account for the different range of definition. This then
gives a `bouncing' packet solution
\begin{equation}
 \tilde{\psi}_{(0)}(x,t) = \left\{ \begin{array}{ll}
\sqrt{2}\psi_{(0)}(x,t)  & \mbox{for $x\leq 0$} \\
0 &  \mbox{for $x\geq 0$}
\end{array} \right.
\, .
\label{novel_bouncer}
\end{equation}

The `bouncing' wave packet solution of Eqn.~(\ref{bouncing_solution}) which
is valid for arbitrary values of $x_0$ and $p_0$ can be considered in the
limit when $x_0 = 0$ and $p_0 \rightarrow 0$. In that limit, we recover
the form in Eqn.~(\ref{novel_bouncer}) (except for trivial multiplicative
factors of $i$.)

For this special solution, almost all of the relevant expectation values
can be obtained in closed form, and we find, for example, that
\begin{equation}
\langle x \rangle_t = -\frac{2\beta_{t}}{\sqrt{\pi}}
\qquad
\mbox{and}
\qquad
\langle x^2 \rangle_t = \frac{3\beta_{t}^2}{2}
\label{special_x}
\end{equation}
which combine to give
\begin{equation}
\Delta x_t = \frac{\beta_t}{\sqrt{2}} \sqrt{\frac{3\pi -8}{\pi}}
\end{equation}
which increases with time in the same way as for the standard free-particle
solution in Eqn.~(\ref{standard_expectations}).

Using the explicit form for $\tilde{\psi}_{(0)}(x,t)$ in 
Eqn.~(\ref{novel_bouncer}),  and the operator form for 
$\hat{p} = (\hbar/i)(\partial/\partial x$), we also find that
\begin{equation}
\langle p \rangle_t = - \frac{2\hbar}{\beta \sqrt{\pi}}
\left[\frac{t/t_0}{\sqrt{1+(t/t_0)^2}}\right]
\qquad
\mbox{and}
\qquad
\langle p^2 \rangle_t = \frac{3\hbar^2}{2\beta^2}
\,. 
\end{equation}
The first of these expressions is consistent with differentiation of 
the result in Eqn.~(\ref{special_x}), while the second is related to 
the free-particle result in Eqn.~(\ref{higher_p}).  The momentum uncertainty 
is then given by
\begin{equation}
\Delta p_t = \frac{\hbar}{\beta} \sqrt{\frac{3}{2} - \frac{4}{\pi}
\frac{(t/t_0)^2}{[1+(t/t_0)^2]}}
\end{equation}
which actually decreases in time. This effect can be understood crudely 
as being due to the fact that the $t=0$ solution has both positive and
negative momentum components in the range $(-\Delta p_0, +\Delta p_0)$,
and over a time interval of order $t_0$, the positive momentum components
are reflected from the infinite wall so that the momentum distribution
is then more localized to the range $(-p_0,0)$.

The effective force on the particle
due to its interaction with the wall, can be associated with
\begin{equation}
\frac{d\langle p \rangle_t}{dt}
= - \left(\frac{2}{\alpha \sqrt{\pi} t_0}\right)
\frac{1}{(1+(t/t_0)^2)^{3/2}}
\end{equation}
which decreases monotonically with time, with an initial value which
scales as $\Delta p_0/t_0$ which is dimensionally correct.

These results can be combined to give the uncertainty principle product as
\begin{eqnarray}
\Delta x_t \cdot \Delta p_t
& = &  \frac{\hbar}{2} \sqrt{\frac{3(3\pi-8)}{\pi}}
\sqrt{1 + (1-8/3\pi)(t/t_0)^2} 
\label{new_uncertainty_product}\\
& \approx & (0.58\hbar) \sqrt{1+(0.15) (t/t_0)^2} \nonumber
\,.
\end{eqnarray}
This localized solution has an initial uncertainty principle product
which is only slightly larger than the minimum value but for long times
is substantially smaller (about half) than that of the standard Gaussian
free-particle solution in Eqn.~(\ref{basic_uncertainty_product}), since
for $t/t_0>>1$, Eqn.~(\ref{new_uncertainty_product}) reduces to
\begin{equation}
\Delta x_t \cdot \Delta p_t 
\quad \longrightarrow
\quad 
 \frac{\hbar}{2}\left(\frac{3\pi-8}{\pi}\right)
\left(\frac{t}{t_0}\right)
= (0.45)  \frac{\hbar}{2}\left(\frac{t}{t_0}\right)
\, .
\end{equation}
Variations on the momentum-space distribution in 
Eqn.~(\ref{alt_p_gaussian_t}) which
are also odd in $p-p_0$ (higher power solutions of the simple harmonic 
oscillator, for example) can be used to evaluate other generalized
solutions for the free-particle case which can also satisfy the
appropriate boundary conditions (at $x=0$) for the `bouncing' particle 
case and provide additional examples.

\section{Conclusions and discussions}
\label{sec:conclusions}

Using the familiar method-of-images or mirror wavefunction technique,
we have shown how to construct normalizable Gaussian solutions with
arbitrary initial $x_0,p_0$ for the {\it bouncing particle} problem, 
for which many expectation values and related quantities are
calculable in closed form. This example adds another case to 
the limited number of time-dependent wave packet solutions of one-dimensional
quantum mechanical problems with familiar classical analogs.  It also 
provides an explicitly calculable example of the effect of the introduction
of a distant boundary on an explicitly time-dependent solution for a 
quantum system for comparison to more general
discussions. Because of the methods used, not all of the expectation values 
(or related quantities such as the momentum-space wave function or the 
Wigner quasi-probability distribution) can be evaluated as easily, but all
of the even expectation values are readily calculable and many of the others,
such as $\langle x\rangle_t$ can be usefully approximated, especially near 
the time of the classical collision with the wall. A class of non-standard
free-particle Gaussian solutions can also be used to provide special
$p_0=0$ which vanish at $x=0$ and are therefore also solutions for 
the bouncing well case, with interesting long-time behavior.

\vskip 1cm

\noindent {\bf Acknowledgments} 

MB was supported in part by a Research Corporation Cottrell College
Science Award (CC5470) and the National Science Foundation (DUE-0126439).

 \newpage

\clearpage

\end{document}